\documentclass[showpacs,preprintnumbers,english,amsmath,amssymb]{revtex4}
\usepackage[T1]{fontenc}
\usepackage[latin1]{inputenc}
\usepackage{graphicx}
\usepackage{amssymb}
\usepackage{verbatim}
\usepackage{epic,eepic}
\usepackage{babel}
\usepackage{color}
\usepackage{float}

\begin{document}

\title{Spin Quantization in Heavy Ion Collision}

\author{Hua Zheng$^{a)}$, and Aldo Bonasera$^{b,c)}$}
\affiliation{
a)School of Physics and Information Technology, Shaanxi Normal University, Xi'an 710119, China;\\
b)Cyclotron Institute, Texas A\&M University, College Station, TX 77843, USA;\\
c)Laboratori Nazionali del Sud, INFN, via Santa Sofia, 62, 95123 Catania, Italy.}


\begin{abstract}
We analyzed recent experimental data on the disassembly of $^{28}$Si into 7$\alpha$ in terms of a hybrid $\alpha$-cluster model.  We calculated the probability of breaking into several $\alpha$-like fragments for high $l$-spin values for identical and non-identical spin zero nuclei. Resonant energies were found for each $l$-value and compared to the data and other theoretical models. Toroidal-like structures were revealed in coordinate and momentum space when averaging over many events at high $l$. The transition from quantum to classical mechanics is highlighted.
\end{abstract}


\maketitle

\section{Introduction}
Recent experimental data \cite{xiguang19} have shown evidence of resonances in the disassembly of the $^{28}$Si nucleus into 7$\alpha$.  The data were obtained from the collision of a $^{28}$Si beam at 35~MeV/A on a $^{12}$C target, the experiment was performed at the Cyclotron institute, Texas A\&M university.  The authors of \cite{xiguang19} tentatively associated these structures to the population of toroidal high-spin isomers as predicted by a number of theoretical models \cite{wheeler, wong1, wong2, wong3, wong4, wong5, wong6, meng1, Ichikawa1, Ichikawa2, wong7, wong8, wong9, Ikeda1, oertzen1, freer1, enyo1, funaki1, wilkinson1}. In particular the experimental analysis concentrated on the disassembly of the projectile nucleus into $\alpha$-like particles.  The data show to a high degree of confidence some structures at excitation energies 114, 126 and 138 MeV, respectively, close to the predicted toroidal state at 143 MeV \cite{wong7}.  These encouraging results call for more experimental and theoretical efforts to uncover these resonances also for different nuclei, different disassembly routes and as a function of excitation energy.  Due to the dynamics involved in the disassembly, microscopic models such as the Anti-symmetrized Molecular Dynamics (AMD) model \cite{amd} or Constrained Molecular Dynamics (CoMD) model \cite{comd} could be used, but they may become numerically difficult to handle when a large number of events is needed. Furthermore, they may not be able to describe in detail the $\alpha$-like events as selected from the data \cite{xiguang19}.  Hybrid models may help in overcoming numerical problems at the expense of some physical insights \cite{charity10}.

It was observed already in the 1930s that $\alpha$-like nuclei ($^{12}$C, $^{16}$O,...)~\cite{alpha1, alpha2, alpha3, alpha4, alpha5, alpha6} display many properties that can be easily explained by assuming that those nuclei are made of $\alpha$ particles with no internal structure.  Inspired by these many works we implemented a dynamical model where $\alpha$ particles interact through suitable chosen two body forces.  We enforced two main variations with respect to what can be found in the literature. The first is the $\alpha$-$\alpha$ interaction.  For simplicity we used the phenomenological Bass potential (for $A$~=~4) widely used for low energy heavy ion collisions \cite{bass}.  This potential is very attractive at short distances thus the particles strongly overlap, overcoming the Coulomb repulsion. As a second ingredient of the model, we treat $\alpha$ particles as Gaussian distributions with widths given by their radius.  Overlapping particles experience Pauli blocking because of the internal structure of the $\alpha$s.  Thus, we include a repulsive effect due to the increase of the Fermi energy opportunely adjusted to take into account finite size effects \cite{aldoreport}.  With such simple assumptions we are able to reproduce the binding energies of even-even N~=~Z nuclei up to mass 104 with less than 5\% discrepancy to the experimental data.  We do not want to stress much the properties of the model since our main goal is to simulate the dynamics of the disassembly to compare to data at least qualitatively.  Furthermore, we would like to confirm or disprove the existence of exotic unstable shapes using a simple and transparent model and hope to be of guidance for future experiments.

\begin{figure}
\centering
\includegraphics[width=0.5\columnwidth]{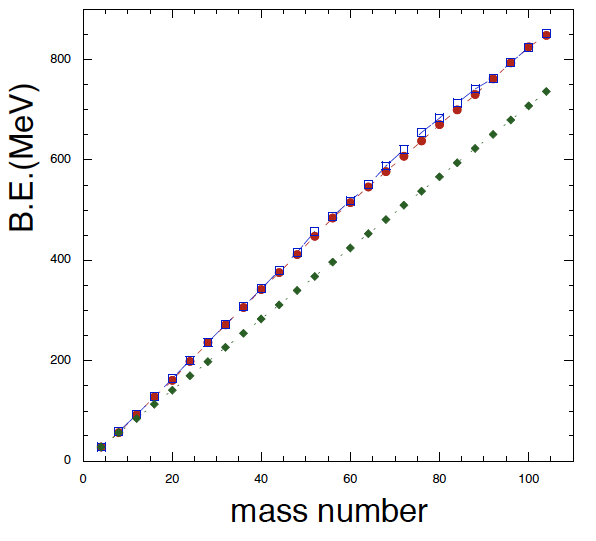}
\caption{(Color online) Binding energy of even-even N=Z nuclei as function of the mass number. Full circles refer to experiments and open squares to the h$\alpha$c model. The full diamonds refer to the number of $\alpha$-particles times the experimental $\alpha$-binding energy.}
\label{f1}
\end{figure} 

We dubbed the model the hybrid $\alpha$-cluster model (h$\alpha$c).  It is a semi-classical model since it includes Pauli blocking effects. In fact the model ground states display strongly overlapping $\alpha$-particles and a strong repulsion due to the increase of the Fermi energy.  It means that ground states cannot be described by $\alpha$-point particles and the nucleons degrees of freedom are essential.  Systems excited by some external probes expand and the $\alpha$-degrees of freedom may become dominant.  Notice that since the repulsion is due to the Pauli blocking and the Coulomb potential, heavy ion collisions using this model can be simulated at energies above the Coulomb barrier up to maybe 80 MeV/A as we discuss below. At higher energies we may need to introduce a suitable collision term, which is a task to be discussed in the future. We introduced another quantum effect in the initial conditions, i.e., we give to the nucleus at time zero a quantized angular momentum $l~=~l_z$ along the $z$-axis.  We assume that the angular momentum is transferred in the collision of $^{28}$Si and $^{12}$C, or $^{28}$Si and $^{28}$Si.  The substantial difference between the two systems is that only even-$l$ values in the entrance channel are allowed in the latter case.  Changing the initial angular momentum revealed a wealth of model features ranging from a first order phase transition of dynamical origin to the formation of short living toroids when averaging over events. Due to its simplicity and numerical affordability, we can make prediction to be tested in future experiments.

\section{The hybrid $\alpha$-cluster model}
In our model $\alpha$-degrees of freedom are treated explicitly while nucleon (protons and neutrons) degrees of freedom are treated implicitly hence the h$\alpha$c acronym.  The interaction between the $\alpha$-particles is given by the Coulomb repulsion (in the monopole--monopole approximation for simplicity) and the nuclear attraction.  The latter is approximated as $V_{\alpha\alpha}=V_{Bass}(A=4)$, i.e., the Bass potential for mass $A=4$ nuclei \cite{bass}.  Coulomb repulsion is not sufficient to prevent a strong overlap among $\alpha$-particles.  Overlapping nuclei increase the repulsion due to the combined action of the Pauli principle and Heisenberg uncertainty principle, in particular the Fermi energy ({\it per} $\alpha$-particle) is given by:
\vspace{6pt}
\begin{equation}
\frac{E_F}{N_\alpha}=4\cdot x_F\cdot \overline{\varepsilon_F}\cdot \overline{\rho}^{2/3}. \label{eq1}
\end{equation}
$\overline{\rho}=\frac{\rho}{\rho_0}$ is the reduced density, $\overline{\varepsilon_F}=\frac{3}{5}\varepsilon_F=21$ MeV is the average kinetic energy of infinite nuclear matter, the factor of four takes into account the fact that we are dealing with $\alpha$ and not nucleons.  For small nuclei corrections are needed to take into account finite size effects, which reduce the Fermi energy thus the parameter $x_F$. In ref. \cite{aldoreport}, Equation 5.5,  such correction was discussed for medium light nuclei resulting in $x_F=0.65$ for $^{8}$Be.  For overlapping $\alpha$-clusters only one nucleon in one $\alpha$ particle is identical to another nucleon in the other $\alpha$.  This parameter takes into account the fact that the Heisenberg uncertainty principle is at play as well for non-identical nucleons \cite{aldopra05}. The overlap between $\alpha$-particles can be described as Gaussian distributions with standard deviation proportional to the $\alpha$-radius $r_\alpha=r_0 4^{1/3}$:
\begin{equation}
\overline{\rho}=2\cdot e^{-\beta(\frac{r}{r_\alpha})^2}. \label{eq2}
\end{equation}
The parameter $\beta=1.22$ is fitted to reproduce the binding energy of $^{12}$C, it is the only free parameter entering the model if we exclude the radius parameter $r_0$. The value of $r_0$ has some consequences regarding the moment of inertia, which we discuss below.  For the purpose of this paper we use $r_0=1.15$ fm, unless otherwise noted, similar to the parameters entering the Bass potential \cite{bass}.  At maximum overlap $\overline{\rho}=2$ and $\frac{\varepsilon_F}{N_\alpha}=86.7$ MeV which is the maximum repulsion in the two body channel to compare to the nuclear attraction $V_{\alpha\alpha}(r=0)=-58$ MeV.  This implies that colliding nuclei will become transparent at beam energies well above the Coulomb barrier, similar to Time Dependent Hartree-Fock calculations \cite{bonche79}.  A suitable collision term may remedy this shortcoming but it is outside the purpose of this work \cite{hua12}.  Equations (\ref{eq1}) and (\ref{eq2}) give the repulsion between particles and we treat it as a classical two-body force.

\begin{figure}
\centering
\includegraphics[width=0.5\columnwidth]{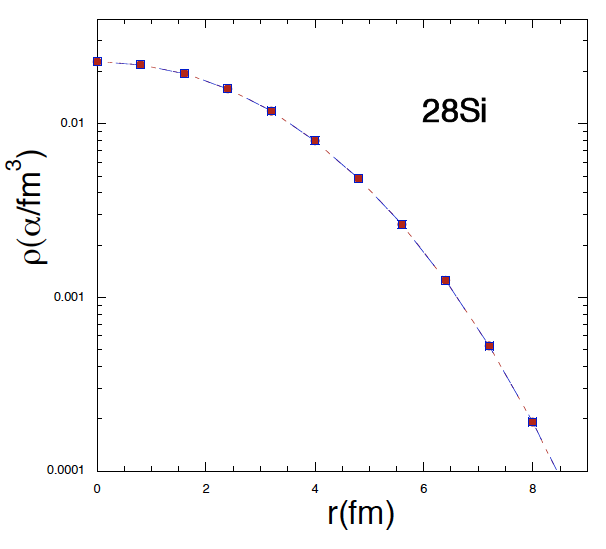}
\caption{(Color online) Ground state density distribution of $^{28}$Si at time $t=150$ fm/c (full symbols) and $t=1500$ fm/c (open symbols). The density is averaged over 50 events.}
\label{f2}
\end{figure} 

The classical Hamilton equations of motion 
\begin{equation}
 \langle \dot {\bf  r_i} \rangle = \frac{\partial H}{\partial \langle {\bf p_i} \rangle}, \quad  \langle \dot {\bf  p_i} \rangle = -\frac{\partial H}{\partial \langle {\bf r_i} \rangle}, \label{comd11}
\end{equation}
for interacting $\alpha$-particles are solved numerically using the O($dt^5$) Runge--Kutta method, $dt=1$ fm/c is a typical time step used in the calculations.  At the highest excitation energies or angular momenta discussed in this paper, the particle velocities become very large thus we implemented relativistic kinematics.  This correction is important but we stress that the description in terms of classical interactions is still valid.  To obtain the nuclear ground states and their binding energies, the equations of motion were solved adding a friction force until a minimum and stable configuration is reached.  The particles position are saved on a file and used as initial positions in dynamical simulations.  To generate events, the initial positions are rotated randomly for each event and/or many different ground states are generated.

In Figure \ref{f1}, we plot the binding energies of $\alpha$-cluster nuclei as function of the mass number $A$.  The free parameter $\beta$ of the model was fixed to reproduce the $^{12}$C binding energy.  This leads to an overestimation of the binding energy of $^{8}$Be of 2.6 MeV: 59.1~MeV theory vs. 56.5 MeV experiment \cite{web}.  This is an important feature since fixing the free parameter to the binding energy of $^{8}$Be would lead to a general underestimation of all the other nuclei.  It implies that the $\alpha$-particles must be more overlapping for heavier nuclei, thus the increase in Fermi energy.  It confirms our discussion above that the correct description of nuclear ground states must be in terms of nucleonic degrees of freedom while $\alpha$-clusters may dominate at lower densities, i.e., in the expansion stage of the nuclear dynamics.  Our hybrid model reproduces the binding energies of nuclei up to mass 104 with an error less than 5\%.  In Figure \ref{f1}, we have included for comparison the contribution to the binding due to the $\alpha$ binding energy, full diamonds.  We notice that changing the value of $r_0$ to 1.26 fm produces a similar agreement to the binding energies with $\beta=1.02$.  Thus, these data are not able to constrain the parameter values to high degree and we will investigate fusion cross sections of even-even N = Z nuclei for further constraints.  

Once the initial conditions are found, we can generate many initial ground states to be used as initial conditions in dynamical calculations.  We treat each $\alpha$-particle as a Gaussian distribution normalized to one of radius (variance) $r_\alpha$.  In Figure \ref{f2}, we plot the density averaged over ensembles at two different times.  Naturally the system is stable and the central density is rather reasonable. The displayed system is $^{28}$Si and we are going to concentrate on this nucleus for the remainder of this paper since it was carefully investigated in ref. \cite{xiguang19}. The calculated binding energy is 236.9 MeV (236.5 MeV from experiments).  An important quantity is the moment of inertia $I$, which can be obtained in our model by opportunely integrating over the density plotted in Figure \ref{f2}. This gives $I=1.3\times 10^5 $ MeV$\cdot$fm$^2$ and $\Psi=\frac{\hbar^2}{2I}=0.15$ MeV, close to the moment of inertia of a sphere of radius $R=1.15A^{1/3}=3.5$ fm and $^{28}$Si mass.  This result should not surprise since the initial configurations are obtained by randomly oriented ground state initial condition and this procedure produces spherical shapes on average.  Notice that increasing the value of $r_0\rightarrow 1.26$ fm gives $\Psi=0.125$ MeV for a sphere, a result that we test briefly below.

\begin{figure}
\centering
\includegraphics[width=0.45\columnwidth]{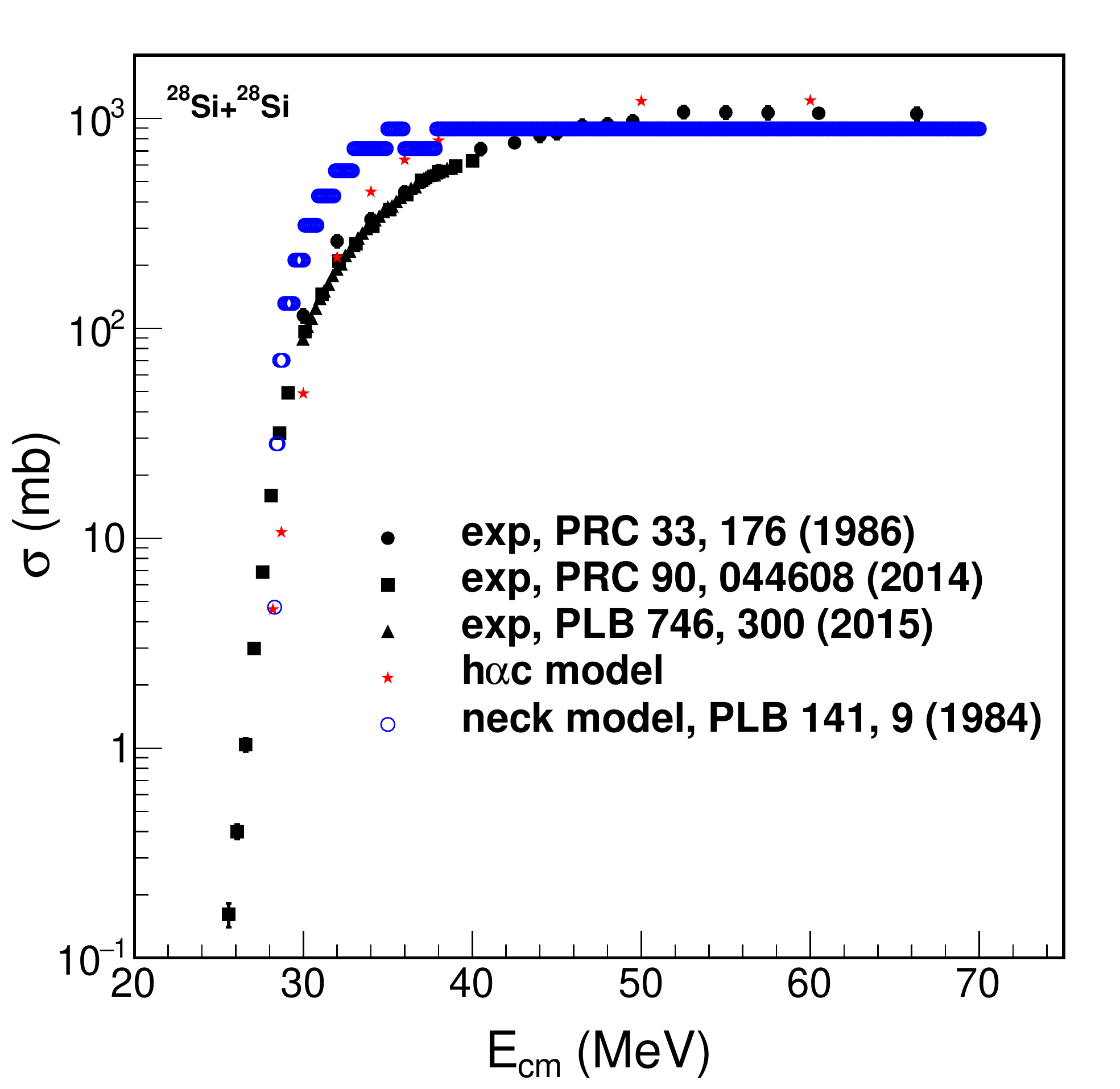}\includegraphics[width=0.45\columnwidth]{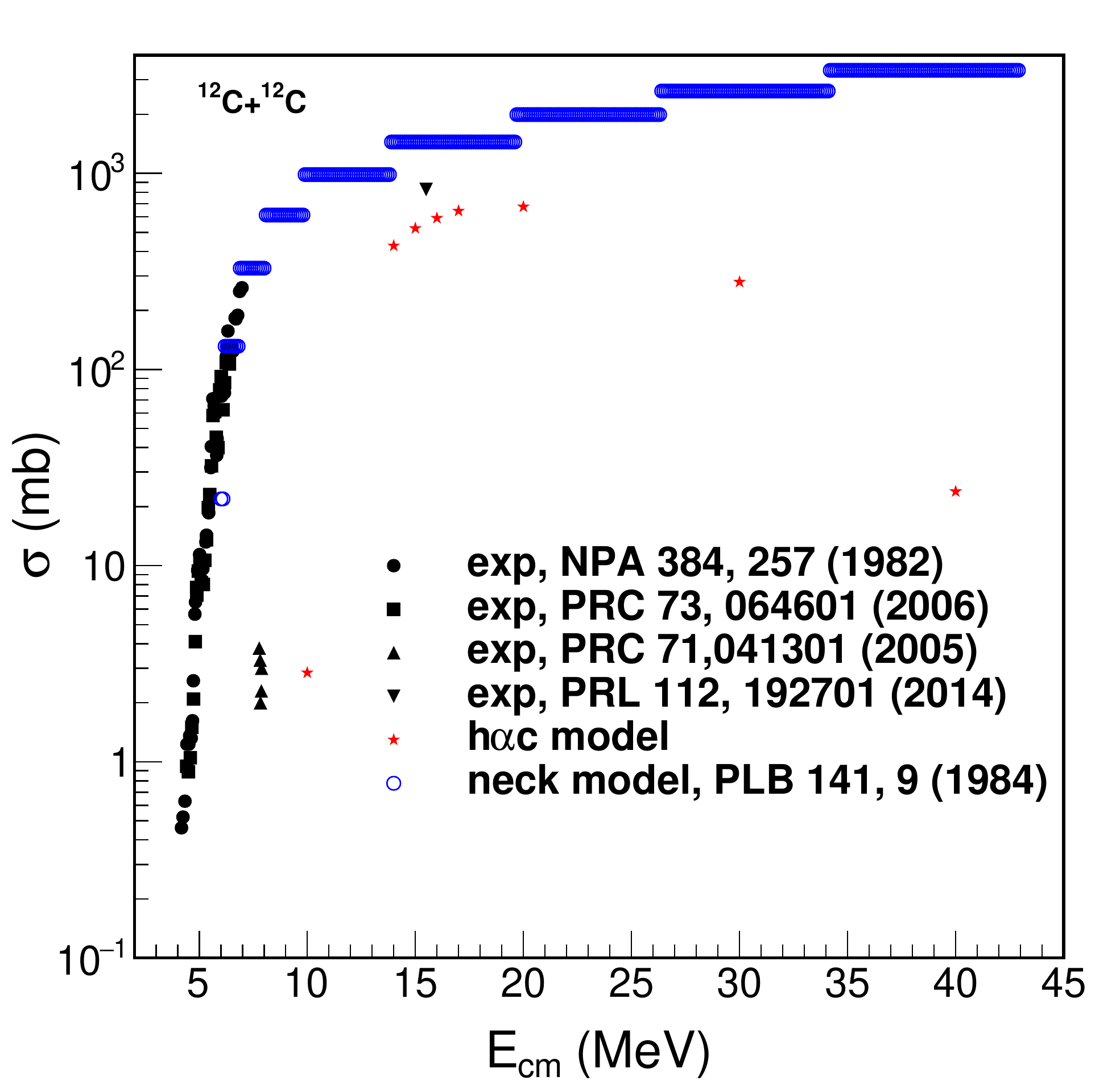}
\caption{(Color online) Fusion cross-section for (left) $^{28}$Si+$^{28}$Si and (right) $^{12}$C+$^{12}$C as function of the center of mass energy.}
\label{f3}
\end{figure} 

\section{Fusion cross sections of identical spin zero nuclei}
The total fusion cross section of the nuclear reaction is
\begin{equation}
\sigma(E_{cm})=\frac{\pi\hbar^2}{2\mu E_{cm}}\sum_{l=0}^\infty (2l+1) \Pi_l = \sum_{l=0}^\infty\sigma_l, \label{eq3}
\end{equation}
where $E_{cm}$ is the reaction energy in the center of mass frame, $\mu$ is the reduced mass of the reaction system and $\Pi_l$ is the fusion probability of the reaction at angular momentum $l$. We simulate the reactions of identical spin zero nuclei, thus only even $l$-values are allowed, $^{28}$Si + $^{28}$Si and $^{12}$C + $^{12}$C at different angular momenta $l$ for a given $E_{cm}$ to obtain the fusion probability $\Pi_l$ with h$\alpha$c model. In Figure \ref{f3}, we plot the fusion cross section of $^{28}$Si + $^{28}$Si and $^{12}$C + $^{12}$C as function of the reaction energy in the center of mass frame. The fusion cross sections calculated from the neck model are also presented for comparison \cite{neckmodel, aldo20}. The h$\alpha$c model can reproduce the experimental cross section data qualitatively. While for $^{12}$C + $^{12}$C at high $E_{cm}$ where there are no experimental data, the difference between neck model and h$\alpha$c model is quite large. Naturally, the model interaction can be improved for a better description of the data, but of course fusion below the barrier needs the inclusion of tunneling \cite{aldo20}.

\section{Rotations and dynamical first order phase transition}
In this section, we will explore the dynamical properties of a $^{28}$Si nucleus rotating along the $z$-axis with initial orbital angular momentum $l=l_z$, in units of $\hbar$. We assume that the orbital angular momenta are transferred through the collision with the $^{28}$Si target nuclei. Due to angular momentum and parity conservation, $l$ must be even in the entrance channel. The amount of angular momentum transferred during the interaction of the two nuclei must be simulated microscopically but this is presently outside the validity of this model.  For simplicity and numerical convenience we will restrict our investigation to even $l$. There are many methods theoretically to give the nucleus an initial angular momentum $l$, a popular one is the cranking model \cite{xiguang19, wong1, wong2, wong3, wong4, wong5, wong6, meng1, Ichikawa1, Ichikawa2, wong7, wong8, wong9}. Since we are dealing with individual particles (7$\alpha$) we will use the following ansatz to give the initial momenta $K_y$, $K_x$ ($\hbar$-units) to particle~$i$:
\begin{equation}
K_y(i)=\frac{x(i)}{r_{xy}^2}l; \quad K_x(i)=-\frac{y(i)}{r_{xy}^2}l. \label{eq4}
\end{equation}

In Equation (\ref{eq4}) $r_{xy}^2=\sum_{i}(x(i)^2+y(i)^2)$ and the sum is extended to all the constituents $\alpha$-particles of the nucleus.  Different events are obtained by different ground states initial positions and we give the initial momenta according to Equation (\ref{eq4}).  Because of the finite number of particles, this will produce classical fluctuations, while the orbital momenta are quantized. This method is more justified when the excitation energy or angular momenta $l$ is larger, i.e., at higher entropies. Of course one should not be surprised that we are utilizing the concept of entropy since our equations of motion are time reversible invariant. Entropy arises from events mixing and averages over phase space.  Notice also that for each event there may be some angular momentum along the $x$ and $y$ directions, see Equation (\ref{eq4}), but on average we have $\langle l_x\rangle=\langle l_y\rangle=0$, this contributes to the initial classical fluctuations. 

\begin{figure}[H]
\centering
\includegraphics[width=0.5\columnwidth]{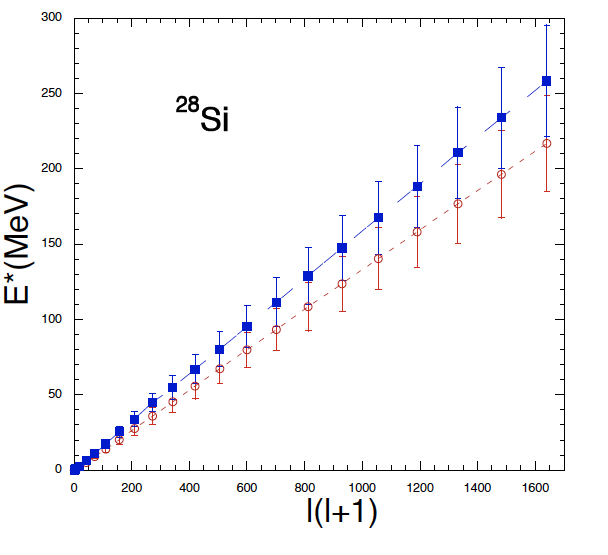}
\caption{(Color online) Excitation energy as function of $l(l+1)$ for $^{28}$Si. Full squares are obtained with the parameter values $r_0=1.15$ fm and $\beta$=1.22, open circles with $r_0=1.26$ fm and $\beta$=1.02. The extracted slopes $\Psi$ are consistent with the values obtained from rotating spheres discussed in the text.}
\label{f4}
\end{figure} 

In Figure \ref{f4}, we plot the excitation energy as function of $l(l+1)$ recovering the familiar linear behavior \cite{ring}. The error bars are given by the variances obtained from event averaging and they are quite small at small $l$ values as expected.  The slope of this plot is proportional to the inverse of the moment of inertia and agrees with our estimate in the previous section for a rigid sphere.  Changing the values of $r_0$ and $\beta$ produces the expected variation, see Figure \ref{f4}.   The moment of inertia is a dynamical quantity since the system expands and breaks into fragments at high $E^*$, thus the obtained value refers to time $t=0$ fm/c.  The calculations give the most probable energies $E^*$ for each value of $l$, the values obtained are reported in Table \ref{t1}.

\begin{table}[H]
	\centering
	\begin{tabular}{|c|c|c|}
		\hline
		$l$ & $l(l+1)$ & $E^*$(MeV)\\
                   \hline
                    0  &     0         & 0.00 \\
  		  2  &     6         & 0.66 (0.58) \\
                    4  &     20       &  2.68(2.2) \\
		  6  &     42       &  6.11 (5.03) \\
		  8  &     72       &  11.0 (8.95) \\
                   10  &    110      & 17.4 (14.0) \\
		 12  &     156     & 25.4 (20.2) \\
                   14 &      210     & 34.0 (27.4) \\
16  &     272         & 44.9 (35.8)\\
18  &     342         & 54.9 (45.3) \\
20  &     420         & 67.1 (55.7) \\
22  &     506         & 80.4 (67.3) \\
24  &     600         & 95.2 (79.8) \\
26 &     702         & 111.0 
 (93.4) \\
28 &     812         & 129.0 (108.0) \\
30  &     930         & 147.0 (124.0) \\
32  &     1056         & 167.0 (140.0) \\
34  &     1190         & 188.0 (158.0) \\
36  &     1332         & 211.0 (177.0) \\
38  &     1482         & 234.0 (196.0) \\
40  &     1640         & 258.0 (217.0) \\
		\hline
	\end{tabular}
	\caption{Most probable energies for each initial angular momentum $l$. In parenthesis the values obtained with $r_0=1.26$ fm and $\beta=1.02$. The experimental values are $E^*$=114, 126 and 138 MeV respectively but for a different system, $^{28}$Si+$^{12}$C \cite{xiguang19} which admits non-even $l$-values (not directly accessible to the experiment).}\label{t1}
\end{table}

\begin{figure}[H]
\centering
\includegraphics[width=0.5\columnwidth]{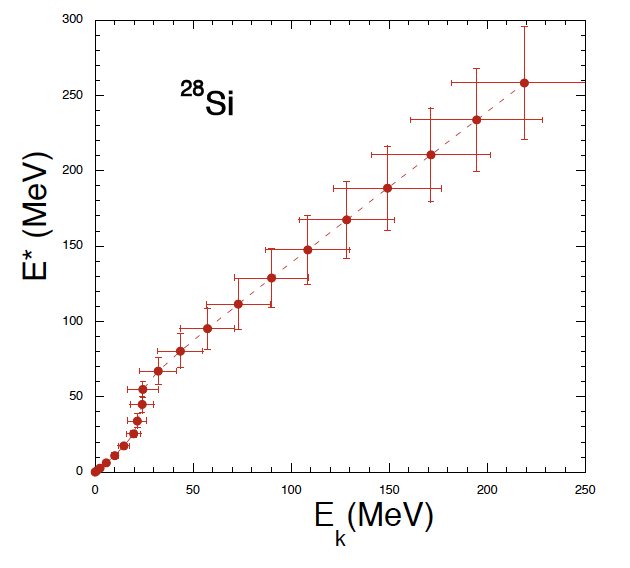}
\caption{(Color online) Excitation energy as function of the kinetic energy of $\alpha$-particles. The steep increase of $E^*$ for $E_k=25$ MeV indicates a first order phase transition. Error bars are the variances of each quantity due to the fluctuations in the initial conditions for fixed $l$.}
\label{f5}
\end{figure} 

An interesting physical quantity is the excitation energy as function of the kinetic energy of the particles. If the system would reach thermal equilibrium, the latter quantity could be related to the temperature.  In Figure \ref{f5}, we plot these quantities and notice the peculiar behavior for $E_k$ near 25 MeV.  The increase in excitation energy for fixed kinetic energy signals the occurrence of a first order phase transition and signals the opening of new channels such as evaporation, `fission' and fragmentation, i.e., the sudden increase of the degrees of freedom of the system, from one nucleus to many fragments. We notice that we get a finite probability of breaking into 7$\alpha$ for $l=16$, and $E^*(7\alpha)= 54$ MeV, which gives the model lowest excitation energy (in 1000 events) when breaking into 7$\alpha$. The question remains if the transition is of thermal origin.  A signature of thermal equilibrium is to observe the same features for each coordinate. We know that the nucleus is rotating along the z-axis, thus we expect the momenta along the same axis to be small, zero on average, see Equation (\ref{eq4}).  To reach equilibrium, energy must be transferred from the other directions and this may be impossible for high $l$-values (and angular momentum conservation). From the large `error bars' (i.e., variances due to the initial conditions) in Figure \ref{f5} we may guess that the system becomes more and more chaotic but that does not prove that thermal equilibrium is reached.  A definite answer to this problem may be obtained by repeating the plot as function of the kinetic energy along the $z$-axis, $E_{kz}$.  If the system reaches thermal equilibrium multiplying $E_{kz}$ by a factor of 3 should reproduce Figure \ref{f5}.

\begin{figure}[H]
\centering
\includegraphics[width=0.5\columnwidth]{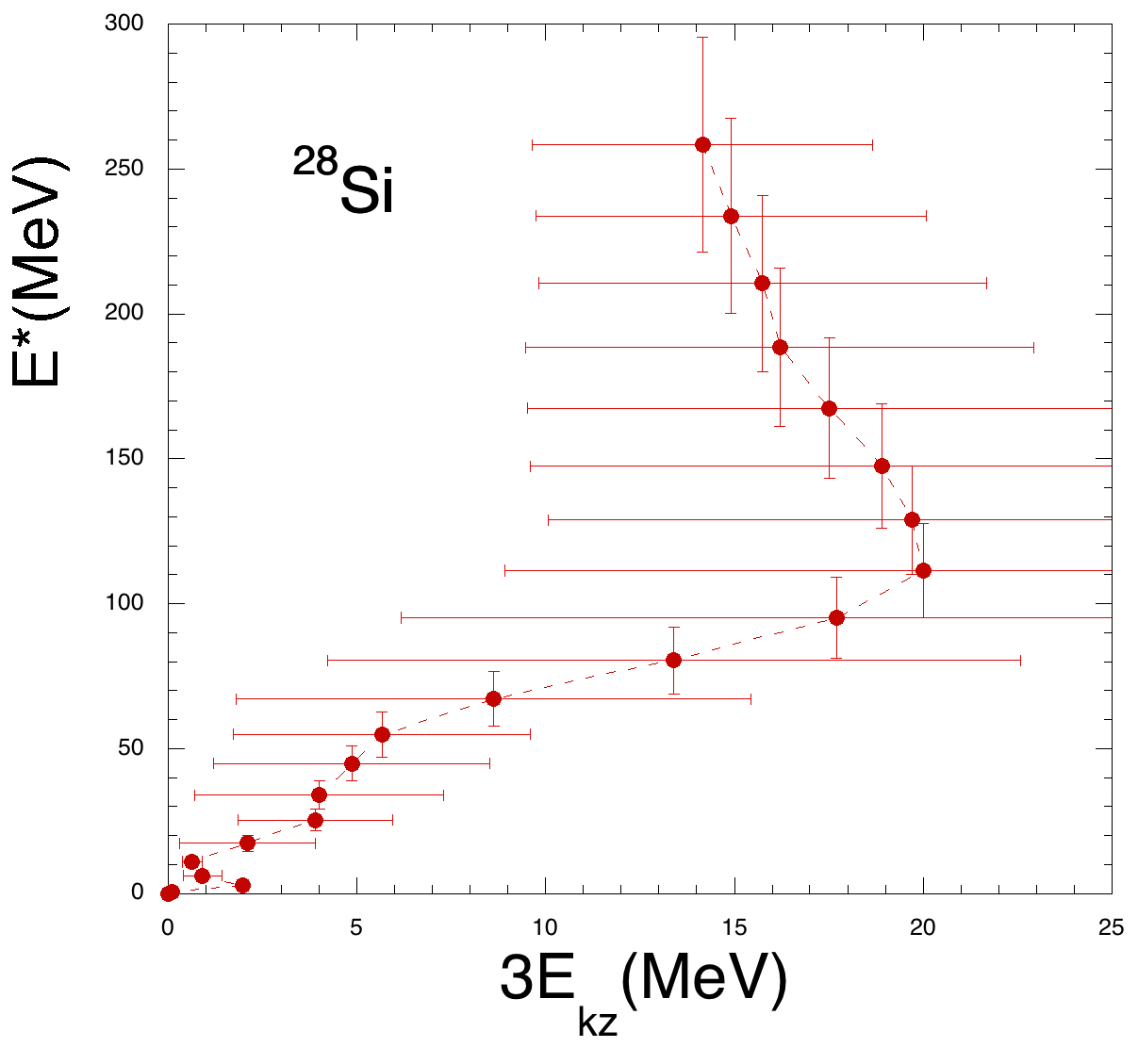}
\caption{(Color online) Excitation energy as function of the kinetic energy (times 3) along the rotation $z$-axis, compare to fig. \ref{f5}.  It indicates an apparent classical maximum `temperature' ($T=2E_{kz}\approx 13$ MeV) the system may sustain. The phase transition is dynamical and due to the opening of new degrees of freedom (evaporation, `fission' and multifragmentation). Toroidal-like structures appear above the phase transition.}
\label{f6}
\end{figure} 

In Figure \ref{f6}, we plot $E^*$ as function of $3\times E_{kz}$, compare to Figure \ref{f5}.  At low energy we observe an increase of $E^*$ up to about 25 MeV where the phase transition occurs.  Higher $l$-values or $E^*$ values do not produce an increase in $E_{kz}$ but just an increase in the variances.  It means that even if we increase the excitation energy the system does not have enough time to transfer kinetic energy from the $x$-$y$ plane to the $z$-direction, i.e., to reach thermalization. This can be interpreted as an apparent maximum temperature that the system may sustain. Such behavior could be compared to the Lyapunov exponent of an expanding system as discussed in ref. \cite{aldoprl95}, it proves that the phase transition is of dynamical origin. These findings could be experimentally tested \cite{xiguang19}.

\begin{figure}[H]
\centering
\includegraphics[width=0.5\columnwidth]{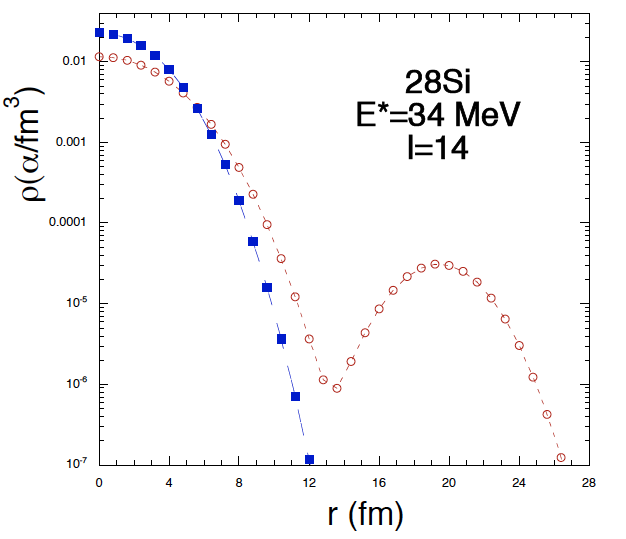}
\caption{(Color online) Density distribution of $^{28}$Si rotating along the $z$-axis. The full symbols are obtained at $t=150$ fm/c and the open ones at $t=1500$ fm/c, compare to figure \ref{f2}. The small density increase at large $r$ is due to $\alpha$-particle evaporation in some events. For this plot, 50 events were generated.}
\label{f7}
\end{figure} 

In Figure \ref{f7}, we plot the density distribution at two different times $t=150$ fm/c and 1500 fm/c at $E^*=34$ MeV, i.e., in the region of the phase transition.  The bump that we observe at later times is due to the escape (evaporation) of one $\alpha$-particle in some events thus doubling the number of degrees of freedom. 

\begin{figure}[H]
\centering
\includegraphics[width=0.4\columnwidth]{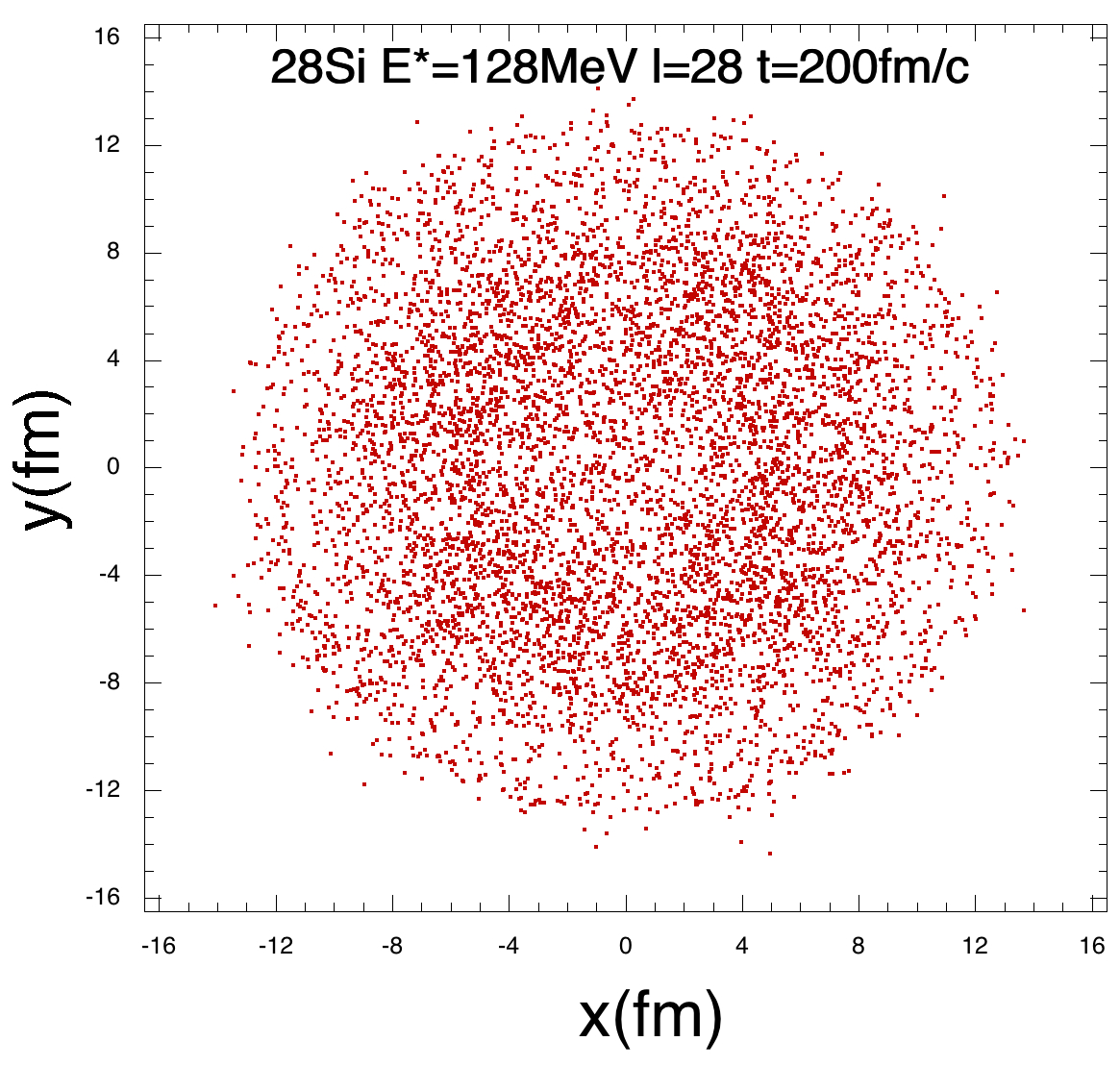}\includegraphics[width=0.45\columnwidth]{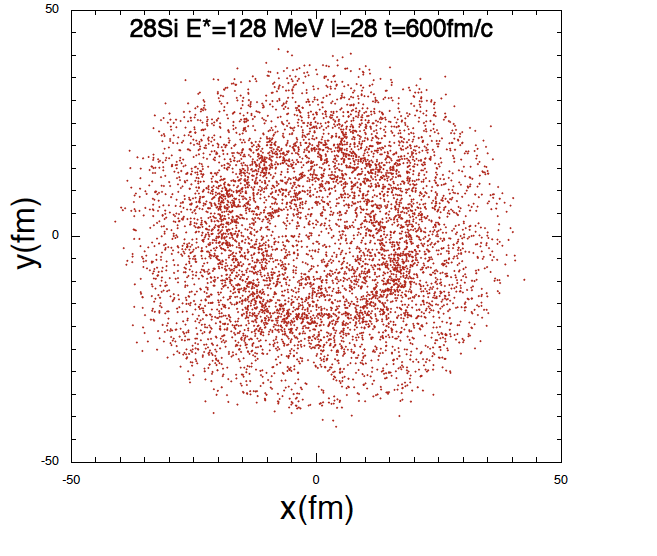}
\caption{(Color online) Time evolution of the disassembly of $^{28}$Si in the rotation plane. 1000 events are generated, only events breaking into 7$\alpha$ are plotted.}
\label{f8}
\end{figure} 

There have been suggestions that highly rotating nuclei may display toroidal shapes \cite{wheeler, wong1, wong2, wong3, wong4, wong5, wong6, meng1, Ichikawa1, Ichikawa2, wong7, wong8, wong9, Ikeda1, oertzen1, freer1, enyo1, funaki1}r and this was also the focus of the experimental investigation in ref. \cite{xiguang19}.  Our model allows us to study the shape evolution as function of time for given $l$ or $E^*$. In \mbox{Figure \ref{f8},} we plot the coordinates of each $\alpha$-particle in the $x$-$y$ plane at two different times, 200 fm/c and 600~fm/c.  Notice that only the events breaking into 7$\alpha$ are included in the plot. We use a simple algorithm to recognize the fragments, i.e., we assume that two particles belong to the same fragment if their relative distance is less than 5 fm.  This may not be the best approach to recognize fragments at earlier times but it is of little importance since we follow the expansion for very long times.  Using this algorithm we can easily estimate the probability of decays into all possible channels allowed by dynamics.

In Figure \ref{f8}, we see that at 200 fm/c matter is missing at the center during the expansion, left panel. At a later time, more $\alpha$-particles are recognized and a toroidal shape is observed, notice the change of scales.  This is due to the combined effect of the angular momentum and averaging over events.

\begin{figure}[H]
\centering
\includegraphics[width=0.46\columnwidth]{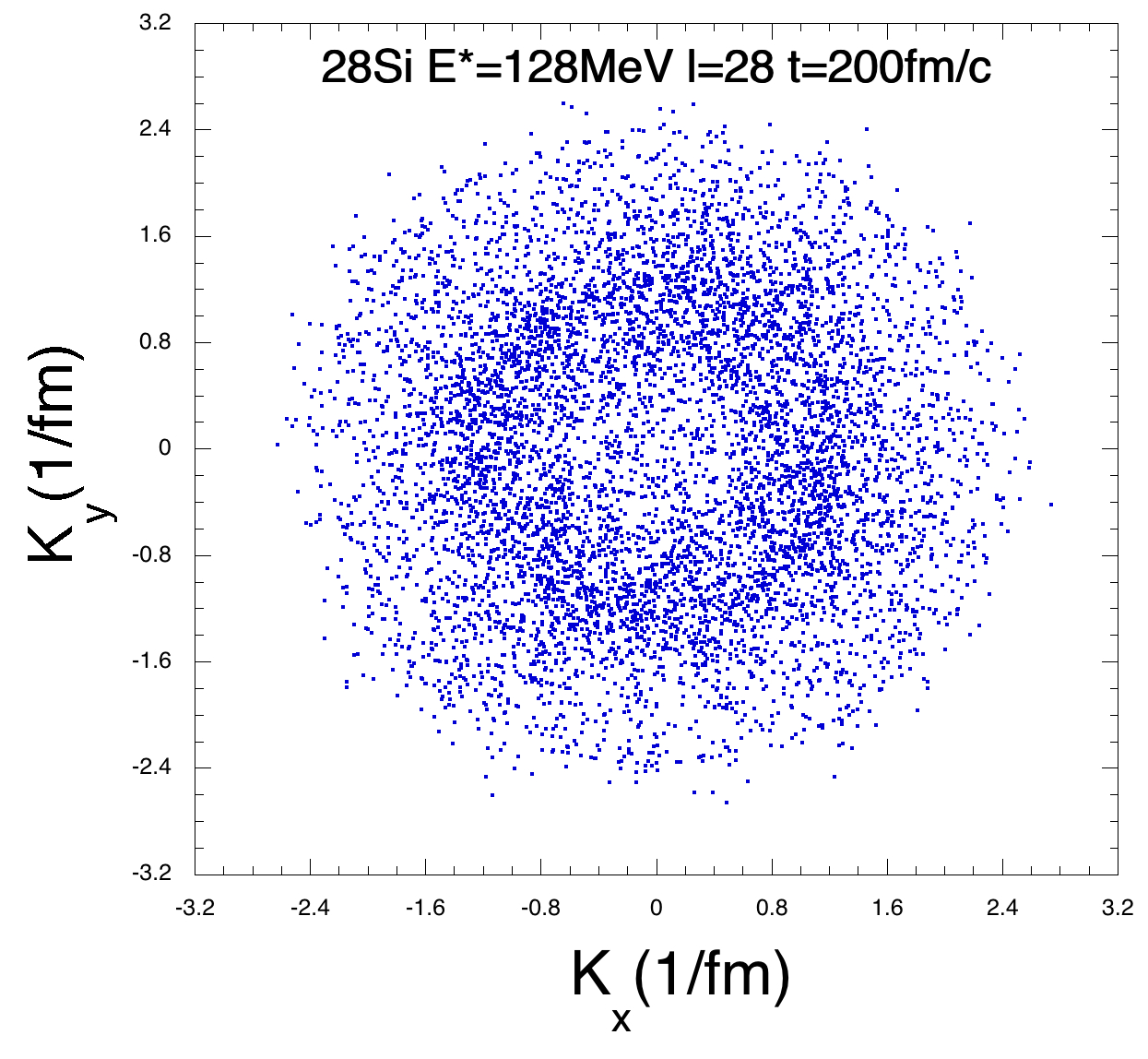}\includegraphics[width=0.45\columnwidth]{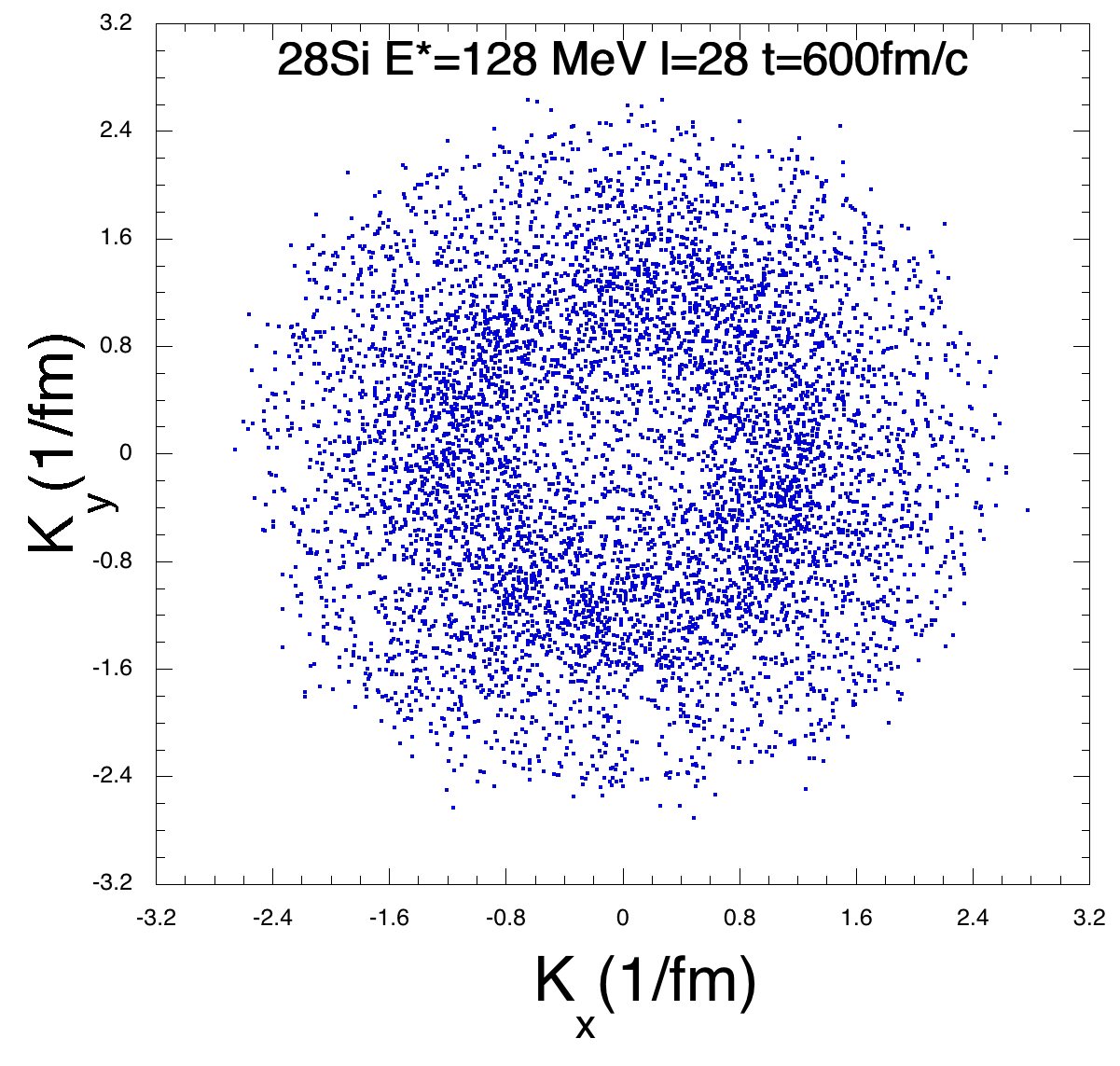}
\caption{(Color online) Same as figure \ref{f8} in momentum space.  Since 7$\alpha$ events only are plotted, fewer events have been recognized at shorter times.}
\label{f9}
\end{figure} 

We can repeat the same considerations in momentum space, see Figure \ref{f9}. At later times the system expands under the influence of Coulomb only and this explains why little expansion is seen in momentum space at the two different times, compare to Figure \ref{f8}.

\section{Cross section estimate}
It is instructive to give an estimate of the cross-section in order to get some deeper insight into the process and be of guidance and stimulus to more experimental and theoretical investigations \cite{xiguang19}. The cross section is given in Equation (\ref{eq3}). We consider the reaction $^{28}$Si on $^{12}$C at 35 MeV/A \cite{xiguang19}, thus $E_{cm}$ = 294 MeV. In order to extend the sum to even $l$-values only, we assume that the angular momenta are transferred in symmetric \mbox{Si + Si} collisions.  $T_l$ gives the probability that in the collision a certain angular momentum $l$ and/or excitation energy $E^*$ is transferred to the $^{28}$Si, see Table \ref{t1}.  This is the part missing in the calculation and we will estimate it from the available phase space in the reaction. We assume that the maximum excitation energy $E_M$ that can be transferred to the Si is proportional to its mass, thus:
\begin{equation}
E_M=\frac{28}{40}E_{cm}=206 \hspace{0.2cm}\text{MeV}. \label{eq5}
\end{equation}
To get this excitation energy in Si + Si we estimate a beam energy E/A~=~29.4 MeV and this could be an interesting experiment to confirm our findings and widen the results of ref. \cite{xiguang19} and also to investigate the angular momentum transfer to each nucleus during the reaction.

Our simple ansatz for the available phase space is:
\begin{equation}
T_l=e^{-\frac{E^*}{E_M-E^*}}. \label{eq6}
\end{equation}
$T_l\rightarrow 0$ if the excitation energy $E^*>E_M$.  This crude approximation guarantees that the cross section vanishes at large excitation energies or large $l$-values. In particular from Table \ref{t1} we expect $l$-values up to 34, while for the set with $r_0=1.26$ fm we expect a contribution up to $l=38$.  More $l$-values included in Equation  (\ref{eq3}) produce larger cross sections, but these cannot be constrained by the data of ref. \cite{xiguang19} since other exit channels including free fermions ($p$, $t$, $^{3}$He etc.) or other fragments are not in the model and these channels may reduce the probabilities especially at high excitation energies.  

\begin{figure}[H]
\centering
\includegraphics[width=0.5\columnwidth]{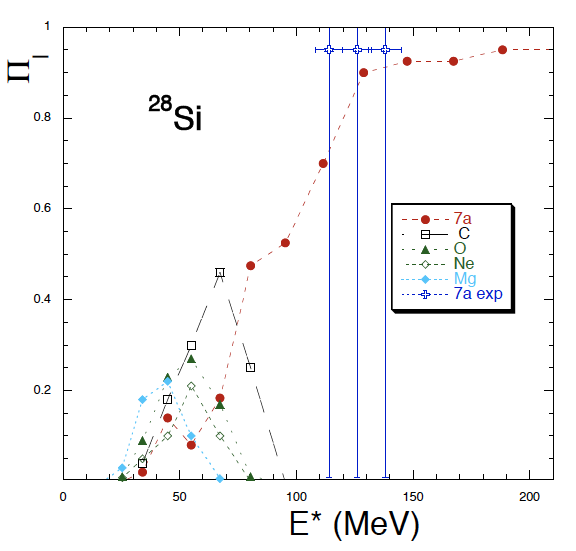}
\caption{(Color online) Decay probability vs excitation energy for $^{28}$Si. Different symbols indicate the largest fragment in the decay. The full circles include decays into 7$\alpha$ and any other combination with one or more $^{8}$Be.  We have included for reference the experimental results \cite{xiguang19} with large error bars since the probabilities are not known. The experimental error on $E^*$ was estimated about 5\% of its value.}
\label{f10}
\end{figure} 

The h$\alpha$c model provides the probability $\Pi_l$ for the system to break into different open channels for a given $l$-value using the simple fragment recognizing algorithm discussed above.  This quantity is plotted in Figure \ref{f10} for the channels as indicated in the inset.  We have included in the 7$\alpha$ channel events where one or more $^{8}$Be are produced, this is to simulate the experimental data where those events are implicitly included; however, recall that the model binding energy of $^{8}$Be is overestimated.  For reference we have included the experimental points from ref. \cite{xiguang19} with huge error bars just to indicate the position of the resonances.  We have seen in the Figures \ref{f8} and \ref{f9} that those values correspond indeed to short living toroidal states. These results can be compared to the toroidal shell model results reported in ref. \cite{xiguang19}, table II.  Notice that the two models differ on the way the orbital angular momenta is given to the system. The 7$\alpha$ channel dominates at high $E^*$ values, while lower resonances are dominated by events where at least one large fragment is present. The larger is the heaviest fragment the smaller is the resonant energy.  This is qualitatively similar to what was observed in the data \cite{xiguang19}.  We stress the fact that all the possible $\alpha$-decay channels can be estimated in the present model. To include channels where nucleons or other fragments are produced we could couple the h$\alpha$c model to a statistical one \cite{charity10}.  These competing channels will decrease the probability that only $\alpha$-channels are relevant, thus we expect that our cross sections are overestimated.

The experiment \cite{xiguang19} provided the reaction cross section per unit energy and different $\alpha$-decay channel.  The h$\alpha$c model gives the energy distribution for each channel whose central values $E^*$ are reported in Table \ref{t1}, and the variances $\Sigma_l$ are reported in the \mbox{Figures \ref{f3}--\ref{f5}.}  We notice that when selecting particular channels (i.e., 7$\alpha$), the most probable energies $E^*$ and their variances change, see Figure \ref{f10}, and we use these values in the following calculations.  We assume that the energy distribution for each $l$-value is given by the Gaussian distribution normalized to one, $g_l(E, E^*, \Sigma_l)$, with units 1/MeV. Thus, we write the differential cross section as:
\begin{equation}
\frac{d\sigma}{dE}=\sum_{l=0}^\infty \sigma_l g_l(E, E^*, \Sigma_l).\label{eq7}
\end{equation}

\begin{figure}[H]
\centering
\includegraphics[width=0.5\columnwidth]{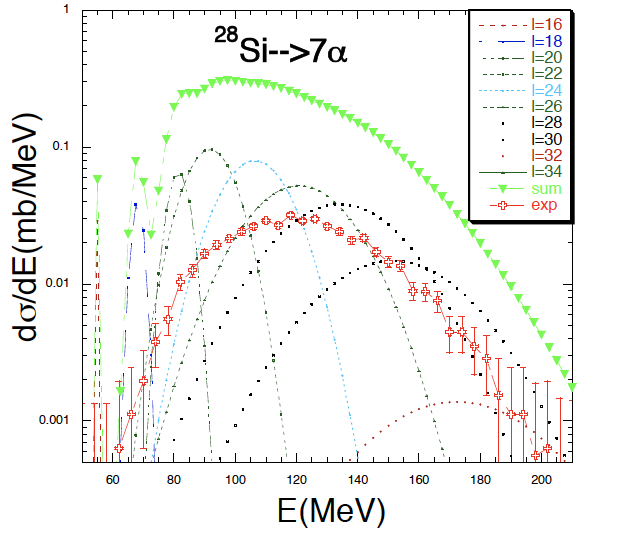}
\caption{(Color online) Differential cross-section as function of the excitation energy for the disassembly of $^{28}$Si into 7$\alpha$.  Contributions for different $l$-values are also included, see inset. Experimental error bars are 4 MeV on the excitation energy and we used the standard error on the differential cross-section. In this plot, `bumps' in the experimental distributions can be noticed at 170 MeV and 190 MeV, not discussed in \cite{xiguang19} because they are in a region of low statistics.  Experiments with higher statistics may confirm or disprove resonances in this energy region, see table \ref{t1}.}
\label{f11}
\end{figure} 

In Figure \ref{f11}, we plot the differential cross section for the 7$\alpha$ channel only.  Different $l$-values contributions are included and indicated in the inset.  The estimated cross-section is generally above the experimental one given by the open crosses. The integrated cross section is a factor 8 above the data (1.9 mb) partly due to the many different open decay channels not included in the model. Notwithstanding these differences there are some interesting features to notice.  The model gives distinguishable bumps at low energies and $l$~=~16--22.  The lowest $l$-values are in a region where experimental values have large error bars.  The experimental cross section starts from 62 MeV (55 MeV in the h$\alpha$c model)~\cite{wada04, meirong21}.  Another feature worth noticing is the increase of the widths for increasing $l$-values, see also Figure \ref{f4}. When those widths become very large, different $l$-values distributions overlap and they cannot be distinguished anymore.  This signals the crossing into classical dominated dynamics where using the impact parameter becomes a good approximation. Our results suggest that repeating the experiment say at higher beam energies may not reveal more values of $E^*$ because they overlap with nearby $l$-values. Lower beam energies may reveal the wealth of distinguishable $E^*$ as in Figure \ref{f11}.  Of course a crucial point would be to have a detector with better granularity and coverage. If not just even $l$-values are admitted in the entrance channel (i.e., Si + C collisions) may open different scenarios. On the same footing experiments using a $^{29}$Si on an identical target would also be interesting to confirm these findings.  The $^{29}$Si beam would have the problem that neutrons are emitted and those are usually difficult to detect in coincidence with 7$\alpha$.

Interesting consequences of our model can be derived from a simple inspection of Equations (\ref{eq4})--(\ref{eq7}) and the oscillations displayed in Figure \ref{f11}.  Indeed similar features have been investigated in the fusion of two heavy ions below and above the Coulomb barrier~\cite{Montagnoli15, aldo20, esbensen12}.  Equation (\ref{eq7}) suggests some simple scaling of the cross section by defining the dimensionless quantity:
\begin{equation}
\xi =\frac{2\mu E E_{cm}}{\pi \hbar^2} \frac{d\sigma (E)}{dE}. \label{eq8}
\end{equation}
In the equation above, the $E_{cm}$ term is important when comparing the same system at two different beam energies.  It does not guarantee overall scaling since at different beam energies, different $l$-values maybe relevant and we expect variations on the tail of the distributions.  Similar to fusion reactions \cite{Montagnoli15}, Equation~(\ref{eq8}) gives the `energy-weighted excitation function' (EWEF).  Notice that in low energy fusion reactions it could be convenient to replace Equation (\ref{eq8}) with \cite{Montagnoli15, esbensen12}: $\xi=\frac{E_{cm}}{\sigma_R}\frac{d\sigma(E_{cm})}{dE_{cm}}$ , where $\sigma_R$ is the reaction cross-section.

\begin{figure}[H]
\centering
\includegraphics[width=0.5\columnwidth]{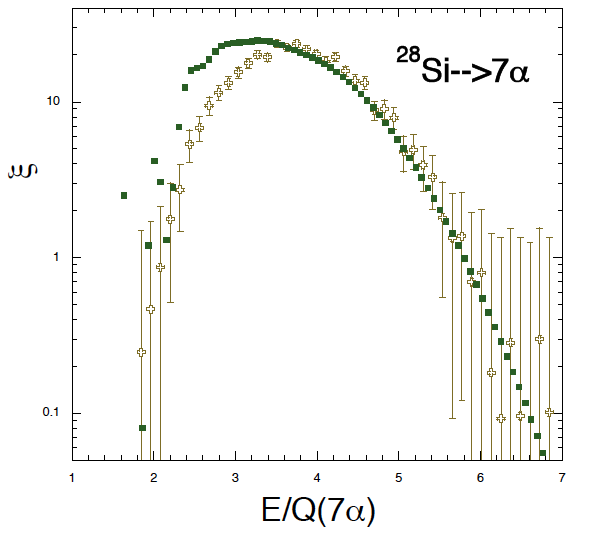}
\caption{(Color online) Scaled differential cross section vs scaled excitation energy, see text. The model results have been divided by a factor 8 to match the data.}
\label{f12}
\end{figure}

Since many exit channels are available in fragmentation reactions, it is useful to define the dimensionless excitation energy as:
\begin{equation}
\overline{E}=\frac{E}{Q}. \label{eq9}
\end{equation}
where the $Q$-value depends on the exit channel, in ref. \cite{xiguang19} the decay of $^{28}$Si in 7$\alpha$ was analyzed in detail and $Q(7\alpha)=33.6$ MeV.  In Figure \ref{f12}, we compare the experimental dimensionless quantities \cite{xiguang19}, open crosses, to the h$\alpha$c model (full squares).  The model calculations have been divided by a factor 8 to take into account the difference to the data in the total cross-section \cite{xiguang19}, compare to Figure \ref{f11}. Standard errors have been included to the data to indicate the region of low statistics, thus of particular interest is the region $2.4<\overline{E}<5.2$.  Figure \ref{f12} does not add much to Figure \ref{f11} but it will become a more interesting observable when data at different beam energies and different combinations of projectile and target will be available.

Oscillations in the EWEF can be better displayed by defining its first and second~derivative:
\begin{equation}
D(\overline{E})=\frac{d\xi}{d\overline{E}}, \quad B(\overline{E})=\frac{d^2\xi}{d\overline{E}^2}.\label{eq10}
\end{equation}

\begin{figure}[H]
\centering
\includegraphics[width=0.5\columnwidth]{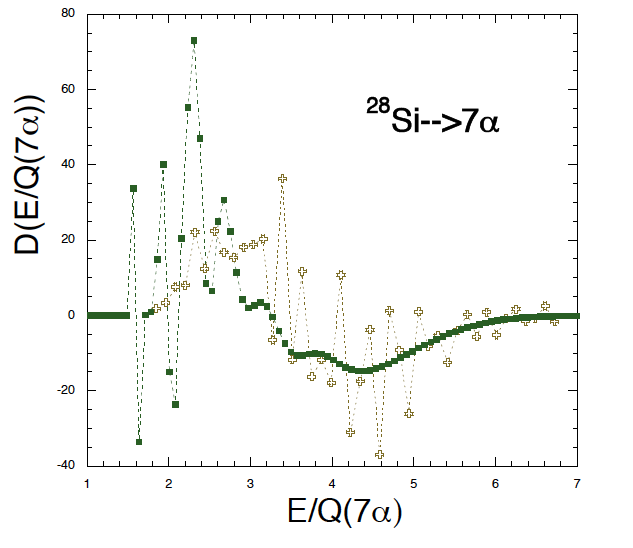}
\caption{(Color online) First derivative of the energy-weighted excitation function vs scaled excitation energy, see text.  Peaks at $E/Q(7\alpha)=$3.4, 3.6 and 4.1 maybe noticed in the experimental data corresponding to $E^*$=114,122 and 138 MeV respectively, compare to ref. \cite{xiguang19} and table \ref{t1}. Peaks at higher or lower scaled excitation energies are in regions where error bars are large (not reported in the figure for clarity), see fig. \ref{f11}.}
\label{f13}
\end{figure} 

The first derivative of the EWEF is plotted in Figure \ref{f13} as function of the dimensionless excitation energy.  Interesting structures in the energies of interest can be noticed.  In particular the data show peaks at $\overline{E}$ = 3.4, 3.6 and 4.1, corresponding to $E^*$~=~114, 122 and 138~MeV, respectively, very close to the data analysis of ref. \cite{xiguang19}.  More peaks may be seen at $\overline{E}$~=~3.1, 4.5 and 4.7 ($E^*$~=~104, 151 and 158 MeV) but in a region where statistics is rather low thus the need for higher statistics experiments.  As expected, the model calculations display definite peaks especially at low excitation energies where the data is poor thus impossible to compare, see also Table \ref{t1}. The average data trend is rather well reproduced by the model.

\begin{figure}[H]
\centering
\includegraphics[width=0.5\columnwidth]{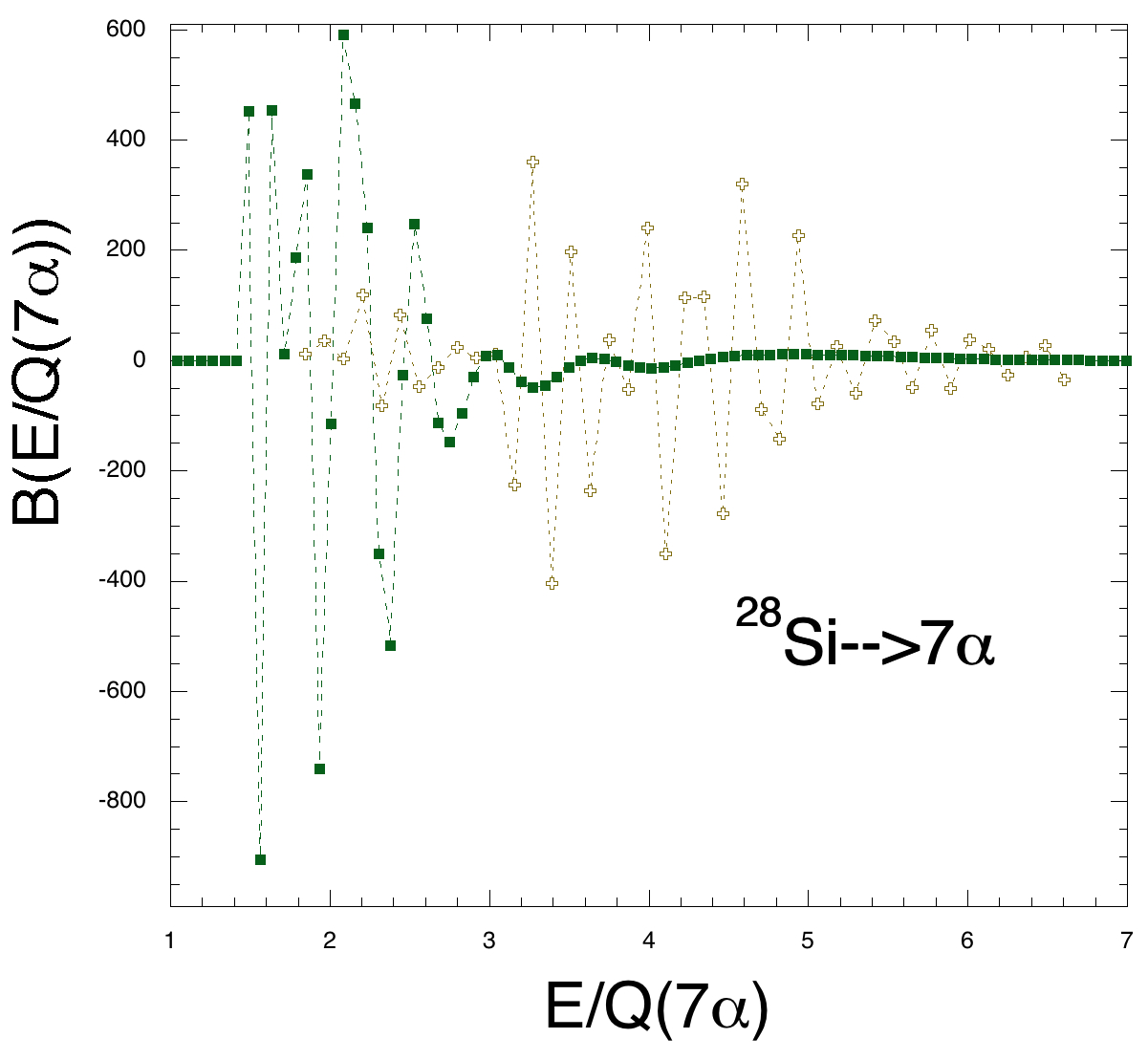}
\caption{(Color online) Second derivative of the energy-weighted excitation function vs scaled excitation energy, see text. Peaks can be noticed as in the figure \ref{f13}. Error bars are not reported in the figure for clarity.}
\label{f14}
\end{figure} 

In Figure \ref{f14}, we repeat our analysis for the second derivative of the EWEF.  In this plot the difference between model and data is more marked. Peaks in the model calculations occur especially for $\overline{E}<$3 while for the data $\overline{E}>$3.  The peaks are consistent to those obtained from the first derivative of the EWEF. 

To complete our comparison to fusion reactions we define the logarithmic derivative as \cite{Montagnoli15, esbensen12}:
\begin{equation}
L(\overline{E})=\frac{D(\overline{E})}{\xi(\overline{E})}. \label{eq11}
\end{equation}

\begin{figure}[H]
\centering
\includegraphics[width=0.5\columnwidth]{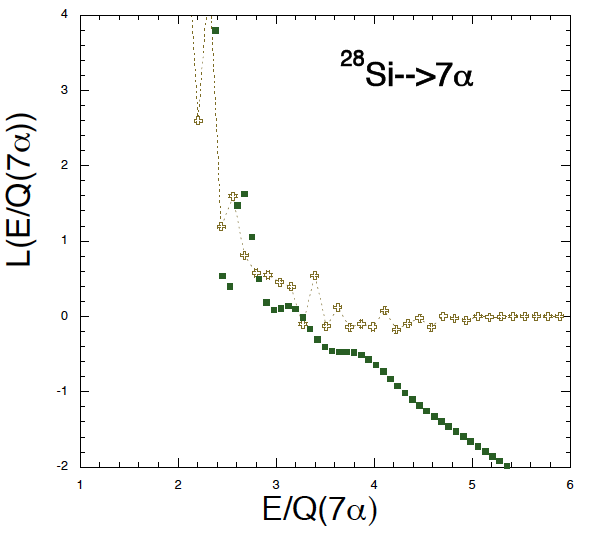}
\caption{(Color online) Logarithmic derivative of the energy-weighted excitation function vs scaled excitation energy, see text. Peaks can be noticed as in figure \ref{f13}.  Error bars are not reported in the figure for clarity.}
\label{f15}
\end{figure} 
This quantity has the obvious property that all constants entering the EWEF and its derivative cancel out, for instance the factor of 8 used to normalize the model to the data.  In Figure \ref{f15}, we plot $L(\overline{E})$ as function of $\overline{E}$ and remark the substantial differences to the previous plots. Peaks in the data remain only for the values reported in ref. \cite{xiguang19}.  The model calculations show large fluctuations in the low energy region and in analogy to the fusion hindrance we can interpret these as a signature of the lowest resonances for this particular decay channel.  Notice however that the model peaks reported in Figure \ref{f11} at \emph{E} = 55 MeV have very low statistics and are not included in this plot (off scale).   If this interpretation is correct we can assume that the $L(\overline{E})$ becomes monotonic at large energies, i.e., in the classical limit.  Unfortunately not much can be inferred from the low energy data since error bars are large, notice however that in the high-energy region of low statistics, the data show a flat behavior thus suggesting that low energy resonances are reachable with more statistics, better detectors and maybe lower beam energy.

\section{Conclusions}
In this work, we have introduced a semi-classical model for nuclei whose constituents are $\alpha$ particles.  In the ground state, the $\alpha$-particles strongly overlap and this gives rise to repulsion due to the increase of the Fermi energy. This means that a proper description of the nuclear ground state must be done in terms of nucleonic degrees of freedom with the possibility that expanding nuclei coalesce into $\alpha$-clusters.  We have applied our hybrid model to the fragmentation results \cite{xiguang19} of $^{28}$Si breaking into even-even N = Z nuclei only.  We have found preferential values of the excitation energies for each $l$-value but also larger and larger variances in the energy distribution due to the fluctuations in the initial conditions, which are classical in origin.  For high $l$-values, the fluctuations become very large and different $l$-distributions overlap signaling the approach to classical mechanics.  We have shown that the spin quantization could be determined experimentally in various different experimental situations by changing the beam energies and the masses of the colliding nuclei including radioactive species.  These experiments require very well performing 4$\pi$ detectors, high statistics and one could take advantage of running the experiments in inverse kinematics.  A dynamical phase transition was also revealed together with the limiting temperature that the system could sustain.  Above the phase transition, toroidal-like shapes are observed when averaging over many events.  These findings may open up a new route of research based on the seminal results of ref.  \cite{xiguang19} and be linked to the oscillations seen in fusion reactions above and below the Coulomb barrier \cite{Montagnoli15, esbensen12}.  We have discussed scaled energy weighted excitation functions and its derivatives.  We have shown how these quantities consistently display interesting features similar to the barrier fluctuations in fusion reactions.  We believe this work may open a new route of investigations to link fusion reaction to deep-inelastic, incomplete fusion and fragmentation.  Well performing 4$\pi$ detectors are needed to improve the energy resolution and granularity and high statistics data are essential.  More theoretical work is finally needed to link the proposed analysis to nuclear fundamental properties.

\begin{acknowledgments} 
We thank prof. J.B. Natowitz, dr. X. G. Cao for providing insight into the experimental data of ref. \cite{xiguang19}. This work was supported in part by the United States Department of Energy under Grant \# DE-FG03-93ER40773 and the NNSA DENA0003841 (CENTAUR),  by the National Natural Science Foundation of China (Grant Nos. 11905120 and 11947416).
\end{acknowledgments}

\end{document}